# *Detection of foraging behavior from accelerometer data using U-Net type convolutional networks*

Mạnh Cường Ngô[a,c #], Raghavendra Selvan[b,d], Outi Tervo[c], Mads Peter Heide-Jørgensen[c], Susanne Ditlevsen[a #]

(a)     *Department of Mathematical Sciences, University of Copenhagen, Universitetsparken 5, 2100 Copenhagen Ø, Denmark*

(b)     *Department of Computer Science, University of Copenhagen, Universitetsparken 1, 2100 Copenhagen Ø, Denmark*

(c)     *Greenland Institute of Natural Resources, Strandgade 91, 2, DK-1401 Copenhagen K, Denmark*

*(d)      Department of Neuroscience, University of Copenhagen, Blegdamsvej 3*, 2200 *Copenhagen Ø, Denmark*

(#) Corresponding authors: nxp418@ku.dk and susanne@math.ku.dk

## Abstract

*Narwhal (Monodon monoceros) is one of the most mysterious marine mammals, due to its isolated habitat in the Arctic region. Tagging is a technology that has the potential to explore the activities of this species, where behavioral information can be collected from instrumented individuals. This includes accelerometer data, diving and acoustic data as well as GPS positioning. An essential element in understanding the ecological role of toothed whales is to characterize their feeding behavior and estimate the amount of food consumption. Buzzes are sounds emitted by toothed whales that are related directly to the foraging behaviors. It is therefore of interest to measure or estimate the rate of buzzing to estimate prey intake. The main goal of this paper is to find a way to detect prey capture attempts directly from accelerometer data, and thus be able to estimate food consumption without the need for the more demanding acoustic data. We develop three automated buzz detection methods based on accelerometer and depth data solely. We use a dataset from five narwhals instrumented in East Greenland in 2018 to train, validate and test a logistic regression model and the state-of-the art machine learning algorithms random forest and deep learning, using the buzzes detected from acoustic data as the ground truth. The deep learning algorithm performed best among the tested methods. We conclude that reliable buzz detectors can be derived from high-frequency-sampling, back-mounted accelerometer tags, thus providing an alternative tool for studies of foraging ecology of marine mammals in their natural environments. We also compare buzz detection with certain movement patterns, such as sudden changes in acceleration (jerks), found in other marine mammal species for estimating prey capture. We find that narwhals do not seem to make big jerks when foraging and conclude that their hunting patterns in that respect differ from other marine mammals.*

## Keywords

*Buzz, accelerometer data, narwhals, East Greenland, convolutional neural network, U-Net, random forest*

## Introduction

The narwhal (*Monodon monoceros*) is a high-Arctic cetacean known for its characteristic tusk (Graham, et al., 2020). It is among the deepest diving cetaceans and can dive to depths of more than 1800 meters (Heide-Jørgensen, 2009). Narwhals dive to forage, and their main prey includes Greenland halibut (*Reinhardtius hippoglossoides*), polar cod (*Boreogadus saida*), capelin (*Ammodytes villosus*) and squids (*Gonatus sp.*) (Heide-Jørgensen, et al., 1994; Laidre & Heide-Jørgensen, 2005). In disphotic and aphotic zones, they need to use acoustics to explore their environment and locate prey, i.e., echolocation, by producing short-duration sounds (clicks) and listening for echoes reflected from surrounding objects (Berta, et al., 2015); and *buzzes*, a series of clicks with short inter-click-interval (below 50 milliseconds) (Blackwell, et al., 2018). The clicks are used for orientation, and buzzes mark the final phase of a potential prey capture event. How frequently they forage and successfully catch a prey is largely unknown due to difficult environmental and logistical conditions in the Arctic that complicate direct studies of prey intake. We have therefore used tagging technologies to collect behavioral data to elucidate the feeding behavior. The movements of five whales were studied during summer in Scoresby Sound of East Greenland in 2018. Facing strong climate changes in the Arctic area, many species in this region are under threat. Understanding foraging behavior of narwhals helps us to understand the conflicts between their food intake and the increasing level of anthropogenic activities, e.g., fisheries and shipping, in the effort to conserve this unique cetacean.

Accelerometer data and acoustic data are widely used in marine mammal science to understand behavior of whales (Nowacek, et al., 2016). Accelerometer data are collected by tri-axial accelerometers that combine two components: the static acceleration due to gravity and the dynamic acceleration due to the motion of the whale, along three axes: surge (longitudinal *X*-axis); sway (*Y*-axis); and heave (vertical *Z*-axis) orientations (Shepard, et al., 2008; Wilson, et al., 2008). Acoustic data are recordings of vibration of medium around the recording devices caused by acoustic radiation (Swanson, 2008). Currently, acoustic data is the best way to estimate potential successful prey capture due to the assumption that a buzz is the sound narwhals make just before attempting to catch their targets. Hence, we assume that it may have some specific movement pattern around the time a buzz occurs. Such a pattern was discovered in harbor seals (*Phoca vitulina*) (Ydesen, et al., 2014), and harbor porpoises (*Phocoena phocoena*) (Wisniewska, et al., 2016), where they made big jerks, i.e., sudden movements, before catching prey. We may not need acoustic data but can use accelerometer data to discover potential prey capture events.

One of the crucial differences between acceleration and acoustic signals in biotelemetry is that acceleration is an easier parameter to collect. Due to the relatively low sampling frequencies of < 500 Hz, data collection can be achieved by less memory and less battery power enabling longer deployments. Accelerometers are therefore also small-in-size enabling their use in various applications for a large range of species. Furthermore, the small size of instruments decreases drags and other negative effects on the individual carrying the tag. Due to the general high sampling rate of acoustic data, acoustics are currently used only in animal-borne archival applications, where the data are stored onboard the instrument. Retrieval of archival instruments can, however, be challenging in habitats such as the polar regions, where ice and extreme seasonal variation of light constrain research to the summer months. Detecting behaviors from acceleration would be an important step for developing satellite-linked biotelemetry applications, that would allow data upload directly from the instrument mounted on the animal. This would in turn extent the temporal and spatial range of behavioral and ecological research.

Analyzing accelerometer and acoustic data to detect behavioral patterns of whales is of considerable interest, e.g., for sperm whale (Fais, et al., 2016) and whale/dolphin (Hillman, et al., 2003). The bulk of these analyses are performed by engineering features from the acquired data and processing them with a suitable prediction algorithm such as logistic regression, support vector machines or random forest. However, feature engineering is a difficult art that requires lots of expert knowledge and is time consuming (Ng, 2015), especially when the data is noisy, vast and if not already well understood. Since the 2000s, the deep learning era has had many breakthroughs from computer vision to natural language understanding, using huge amount of input data without (much) feature engineering like in traditional machine learning approaches (Alsheikh, et al., 2015; Goodfellow, et al., 2016). Convolutional Neural Networks (CNN) are among the most widely used deep learning architectures (Krizhevsky, et al., 2012; Farabet, et al., 2013; Szegedy, et al., 2015; Tompson, et al., 2015). Unlike in the classical machine learning methods, where one needs to design filters by traditional engineering, CNN can "learn" features directly from the data by its huge number of parameters.

Given the huge amount of data acquired from the animal-borne instruments when sampled at high frequency, we have chosen to explore CNN-based methods for developing robust methods to detect buzzes from accelerometer data. Analyzing time series data at multiple resolutions allows capturing useful temporal correlations and renders it suitable for modelling behavior of narwhals. Our baseline machine learning approach is random forest, which has been used extensively in human accelerometer datasets (e.g., see (Kwapisz, et al., 2010; Bayat, et al., 2014)) as well as in animal studies (Shepard, et al., 2008; Wilson, et al., 2008; Wang, 2019). Furthermore, we compare our CNN model with logistic regression. We hypothesize that there exists some hidden movement pattern during or around the buzzes. Therefore, the main goal in this work is to find patterns that might be descriptive signals during buzzing events of narwhals detected using machine learning algorithms, including traditional ones like random forest, advanced ones like deep learning, or logistic regression. Furthermore, we will also investigate if jerks are correlated with buzz events.

## Material and Methods

### Ethics statement

Permission for capturing, handling, and tagging of narwhals was provided by the Government of Greenland (Case ID 2010±035453, document number 429 926). The project was reviewed and approved by the IACUC of the University of Copenhagen (June 17th, 2015). Access and permits to use land facilities in Scoresby Sound were provided by the Government of Greenland. No protected species were sampled.

### Data

Five male narwhals were instrumented with Acousonde™ acoustic and orientation recorders (www.acousonde.com) in Scoresby Sound, East Greenland in 2018 (Figure 1), for details of the tagging methods see (Heide-Jørgensen, et al., 2015; Heide-Jørgensen, et al., (In prep)). Narwhal acoustic signals can reliably be detected using a relatively low sampling rate (Blackwell, et al., 2018) and the deployments in this work used continuous sampling at 25,811 Hz (16 bit-resolution). In addition, a 3D accelerometer sampled the movements of the whales at 100 Hz and a pressure transducer sampled the depth of the whale at 10 Hz. The data were initially collected to analyze the effects of noise from seismic exploration on the natural lives of narwhals in East Greenland. To avoid interference with altered behavior it was decided to only include data collected before the seismic exposure. In addition, we removed the first 24 hours in all five data sets to eliminate a possible influence from the capturing and tagging (Tervo, et al., 2020). Their IDs were 21791, 20158, 20160, 168433, and 168437. The data set is large with a total length of 121.8 hours and 43,841,100 datapoints for the five narwhals, with a total of 2615 buzzes whose total length is 1 hour 31 minutes 56.8 seconds.

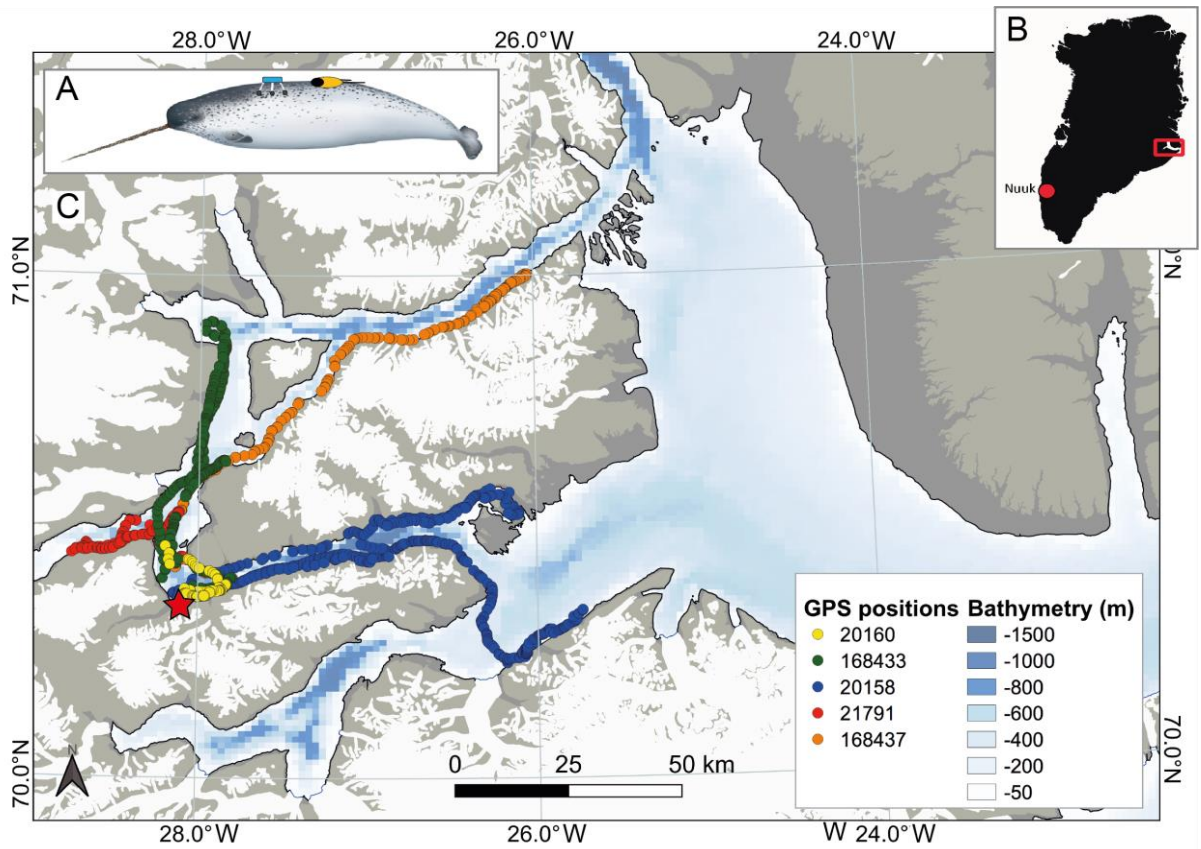

*Figure 1: Scheme of the placement of GPS saddle-back tag (light blue) and Acousonde TM behavioral tag (orange) on a narwhal (A), map of Greenland showing the Scoresby Sound fjord (red box) in East Greenland (B) and a zoomed in map of the study area with the location of the field site, Hjørnedal (marked with a red star), where tagging of narwhals took place (C). The tracks of the five male narwhals used in this study are shown as hourly mean GPS positions. Illustration of a narwhal by Uko Gortner.*

All sound files of 30-minute length from the Acousondes were examined manually by two analysts in MTViewer (a custom-written program for analysis of Acousonde data, W.C. Burgess, personal communication) for continuous click trains produced by the whale. A custom-written buzz detector (Matlab, The MathWorks, Inc., Natick, MA, USA) was used to identify buzzes made by the whales; all positive buzz detections were verified manually by experienced manual analysts. Each file's acoustic signal was reviewed visually, then the analyst listened to the first five seconds of each buzz to estimate the background noise. The data set was too big to listen to in its entirety. The lengths of buzzes vary from 0.4 seconds to 6.7 seconds. The positive rates of buzz labels, i.e., the sum of the lengths of buzzes over the total length of the data, of each whale were: 1.37% for narwhal 21791, 0.73% for narwhal 20158, 1.77% for narwhal 168437, 1.12% for narwhal 20160, and 1.40% for narwhal 168433. Note that these lengths were estimated by an automatic detector, and not as accurate if manually recorded. The buzz resolution was 10 Hz; hence it was expanded to 100 Hz to fit the resolution of the accelerometer data. The true dynamic component of accelerometer data is difficult to extract without gyroscopes or speedometers and is unknown for narwhals, so we used the raw data to let the algorithms explore it directly. We included depth as a feature since the buzz distribution strongly depends on depth (Blackwell, et al., 2018).

We tested whether there was an association between buzzes and jerks. A jerk is defined by the norm of the differences of the acceleration of each axis. We define an RMS jerk to be the root-mean-square (RMS) of the three jerk values over a window of 200 milliseconds, i.e., over a total of 3 × 20 = 60 data points (Ydesen, et al., 2014). The calculation of RMS over a window was used for smoothing the accelerometer data and attenuate the impact of clipping, i.e., when the signal is larger than the detection threshold (Ydesen, et al., 2014). In each window, we defined a buzz to happen if the whale was buzzing at least half of the duration of the window. We defined a big RMS jerk to happen if the jerk's RMS was above pre-defined thresholds defined below. We defined a window to be a positive if there was a big RMS jerk (negative if no big RMS jerk), and a true positive if there was both a big RMS jerk and a buzz, and likewise for false positives and true/false negatives. We calculated the precision and the recall for different thresholds for each whale, where the precision is defined as the ratio of the number of true positives over the sum of true positives and false positives, and the recall is defined as the ratio of the number of true positives over the sum of true positives and false negatives. The thresholds were chosen between 0 and 166,000 mG/s, slightly larger than the maximum value of RMS jerks measured in the data. The thresholds were evaluated in steps of 2000 mG/s, where 1 G = 9.81 m/s$^2$. We calculated precision and recall for instantaneous big RMS jerks (at the same time as the buzz), as well as for delays of 0.2, 0.4, 0.6, 0.8 and 1 second, respectively, to check if the big RMS jerks happen after the buzz.

We defined a dive as a continuous period during which the maximum depth is at least 20 meters, while 10 meters was chosen as the onset and the end of a dive (Figure 2). We chose 10 meters due to possible wrong zero-offsets of the dive (Luque & Fried, 2011). A dive was separated into three phases: the descending phase, the bottom phase, and the ascending phase, apart from the surface phase. The bottom phase was defined as the period at which the whale spent at or below 75% of the maximum depth of the dive (Tervo, et al., 2020). The descent phase was defined as the period between the onset of the dive and the onset of the bottom phase, and the ascent phase was defined as the period between the end of the bottom phase and the end of the dive.

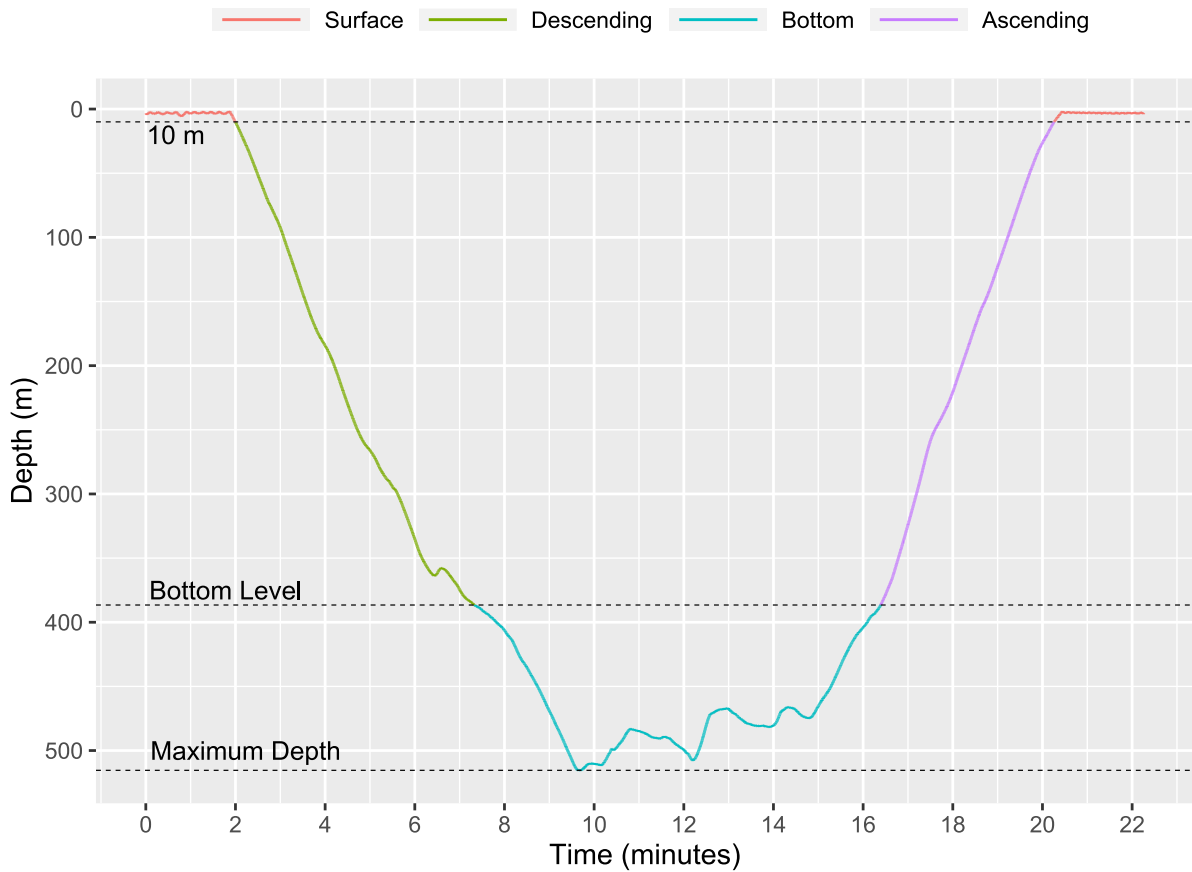

*Figure 2: Example of a dive of the record of narwhal 168433. The dashed line "Bottom Level" is the depth threshold (75% of the maximum depth) for the bottom phase. Red curves, green curve, cyan curve, and violet curve indicate surface, descending, bottom, and ascending phases, respectively.*

The phase of a dive is important for the buzzing activity, since the whales typically forage and buzz at the bottom of the dive, and buzz less during descend and ascend. The categorical feature describing the four dive phases was included by one-hot encoding where each phase is transformed into a binary vector: surface as (1,0,0,0), descending as (0,1,0,0), bottom as (0,0,1,0), and ascending as (0,0,0,1). Thus, we have five features: the three accelerometer axes $A_X, A_Y, A_Z$, the depth, and the diving phase. An example of a subsample of the record of narwhal 21791 shows the features during the bottom phase of a dive together with the response variable of buzzing, where one encodes that a buzz is happening (Figure 3).

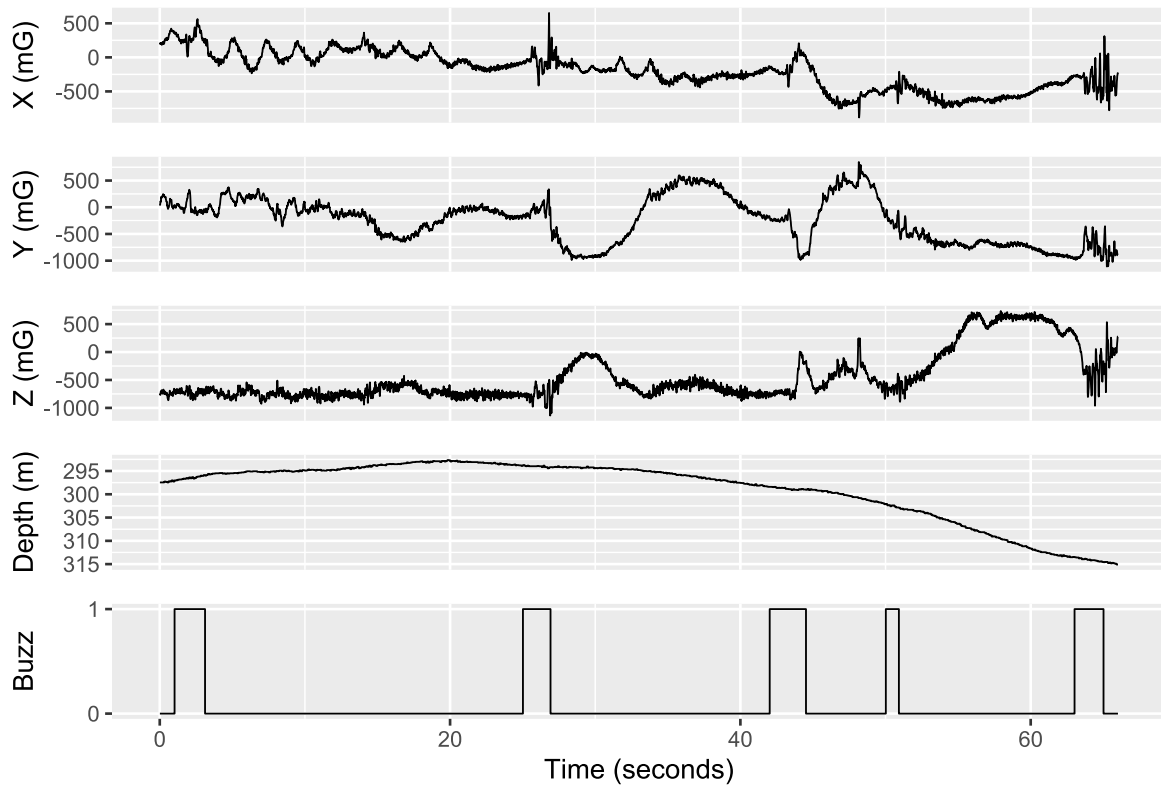

*Figure 3: Example of the record of narwhal 21791 showing the time evolution of the four features and the response variable Buzz, during the bottom phase of a dive. The panels show the 3D accelerometer data, the depth, and the buzzes (1 is presence, and 0 is absence). Note the increased variability in the accelerometer data during some of the buzzes.*

*Supervised U-Net*

Recently, Perslev *et al.* (Perslev, et al., 2019) used a U-Net encoder-decoder architecture, a specific design using CNN as the base (Ronneberger, et al., 2015), for multidimensional time series, called U-Time. The U-Net originally was designed for image segmentation tasks (Ronneberger, et al., 2015). It uses an encoder-decoder type architecture as shown in Figure 4. U-Net encodes input data to feature maps at multiple resolutions, by applying convolution layers followed by downsampling layers (using max-pooling) in the encoder. The sequence of steps in the encoder, also called the contracting path, allow the convolution layers to learn useful features at different resolutions of the data. Then the decoder up-samples such encoded features through an up-sampling layer, then concatenates with the corresponding feature maps from the encoder through skip connections (Drozdzal, et al., 2016). It helps the decoder to have detailed information in the earlier stages from the contracting path, which is lost due to pooling layers in the encoder. Moreover, skip connections make the model easier to optimize in practice (He, et al., 2016). The output of the decoder in a U-Net makes predictions at full resolution in the output data. Perslev *et al.* have shown that their U-Net model, the U-Time, has obtained similar performance as Recurrent Neural Networks (RNN) (Williams, et al., 1986), the default choice for time series data, while RNN is harder to train. Therefore, following their work, we used U-Time architecture as the deep learning model for detecting buzzes from the input data.

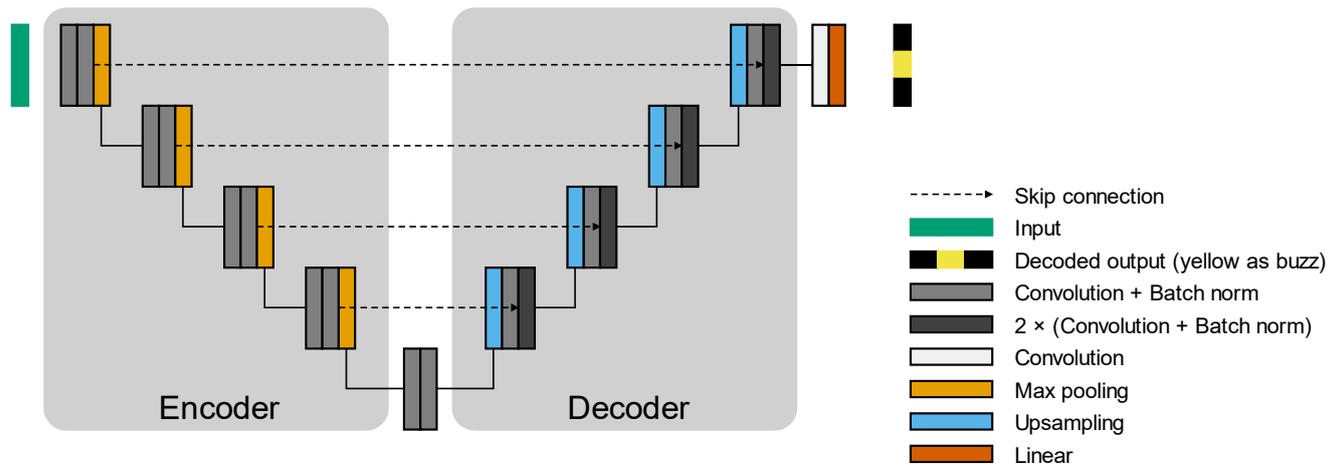

*Figure 4: Structure of a U-Time network. The dashed lines are skip connections. The continuous lines connect the encoder and the decoder. The inputs are time series (accelerometer and depth data), while the yellow interval indicates the duration when buzz happens.*

*U-Net Implementation*

We implemented the U-Time/U-Net model (Ronneberger, et al., 2015; Perslev, et al., 2019) in Python 3.6.9 using PyTorch 1.6.0 (Paszke, et al., 2017) on Google Collaboratory with NVIDIA P100 of 16 GB of RAM (Google, 2020). We used the same architecture as (Perslev, et al., 2019), except that we replaced the last activation function `softmax` with `sigmoid`, since the classification problem is binary.

*Optimization objective*

The dataset is very imbalanced (buzzes occur only rarely in the dataset). This makes the standard machine learning algorithms, including deep learning, perform poorly. Accuracy is the default loss used by most algorithms, where wrongly predicted zeros and ones are penalized the same. However, when most labels are the same (in the dataset, more than 98% of the responses are zeros), the prediction of all being zero (i.e., concentrating on the majority) will lead to a high accuracy, but be uninformative for the problem (Visa & Ralescu, 2005).

We therefore used the Dice loss ($DL$) (Smith, et al., 2020), which is a loss designed for highly imbalanced data, where wrongly predicted ones are penalized more than wrongly predicted zeros, defined by

$$DL = 1 - \frac{2 \sum_{i=1}^{N} p_i g_i}{\sum_{i=1}^{N} p_i + \sum_{i=1}^{N} g_i}$$

where $p_i$ is the predicted probability of a buzz at time $i$, and $g_i$ is the ground truth (the observed buzz or not) at time $i$ and takes values 0 or 1 for $i = 1, \dots, N$.

Let $\alpha = \frac{1}{N} \times \sum_{i=1}^{N} g_i$ be the proportion of ones in the data set, where $0 < \alpha \ll 1$ since the dataset is imbalanced. If the model predicts all 0's, i.e. $p_i = 0$ for all $i = 1, \ldots, N$, then $DL = 1$ because $p_i g_i = 0$ for all $i = 1, \ldots, N$. If the model predicts all 1's, i.e. $p_i = 1$ for all $i = 1, \ldots, N$, then $DL > 1 - 2\alpha \approx 1$ since $\alpha \ll 1$. On the other hand, if the model predicts all correctly, then DL = 0. Therefore, Dice loss penalizes effectively if the model predicts all 0's or all 1's.

*Model selection*

We divided the data set into training, validation and test sets following a ratio of 60:20:20 chronologically for each of the five whales, then combined the training, validation, and test sets from each whale. We also performed *cross validation* on each of the whales, i.e., trained on three whales, validated on one, and used the last one for testing, to evaluate how well the trained model generalized to a new dataset. The Dice loss was computed between the model predictions and the ground truth of the training data. The validation data is used to avoid over-fitting. The difference between the model's prediction and the ground truth was measured by the Dice loss. Stochastic gradient-based optimization algorithms were used on the training data to update the model parameters gradually by iterations, denoted epochs. The validation set was used to select the epoch at which the validation loss, the loss measured on the validation set, was minimal, before the models overfitted. The trained models were then evaluated independently on the test sets to avoid data leakage, a phenomenon where there is some information leakage from validation sets or test sets into training sets.

*Optimization*

We did hyper-parameter search for batch size and the number of convolution filters at the first convolutional layer of the U-Net. Batch size, related to mini-batch gradient descent, is primarily used to smooth the gradients, and can be parallelized. Convolution filters, or filter banks (whose parameters need to be learnt), are used to transform the input to feature maps. The number of hidden units varied between two, four, eight, and sixteen, while batch sizes varied between two, four, eight, and sixteen during our preliminary experiments. We used Adam (Kingma & Ba, 2015) with different learning rates between 0.01, $5\times10^{-3}$, $10^{-3}$, and $5\times10^{-4}$. Smaller learning rates did not help to make the model converge after trial and error. We ran until 301 epochs but did early stopping if there were no improvements after 150 epochs since the best epoch, i.e., the epoch at which the loss function is minimum. We decided to tune the best hyperparameters on only one whale, the data from narwhal 168433, since it was computationally expensive.

*Random forest and logistic regression implementation*

There are no hyper parameters in logistic regression, and therefore no need for a validation set. We therefore divided the data set into training and test sets following a ratio of 80:20 for each of the five whales. We implemented logistic regression (Harrell, 2015) and random forest (Ho, 1995; Breiman, 2001; Boehmke & Greenwell, 2020) as our baseline methods to compare with U-Net. We used the default hyperparameter setting of random forest in Scikit-learn (Pedregosa, et al., 2011), which works well in most cases (Probst, et al., 2019). We implemented random forest with balanced subsample, i.e., for each tree we assigned greater weights to the minority class (here the positive class) based on its bootstrap sample (Chen, et al., 2004). To improve the learning of these models, manual feature extraction should be done. Selecting the right features is a difficult and intricate task, however, these two traditional methods are more robust and easier to interpret (Warmerdam, 2018).

Feature extraction was done for both logistic regression and random forest, following Bayat, et al. (2014). The data were divided into successive windows consisting of 100 consecutive data points, i.e., one second, to compress information into features. Each window shared an overlap of 50 data points with the next window, to assure that no specific pattern was broken due to the edges of these windows (Figure 5). We used twenty-seven features:

- Mean, standard deviation (STD), root mean square (RMS) and MinMax (the difference between maximum and minimum within a window) of accelerometer components $A_X, A_Y, A_Z$ along three axes $X, Y, Z$, as well as the mean depth (13 features).
- STD, RMS and MinMax of the magnitude of the acceleration $A_m = \sqrt{A_X^2 + A_Y^2 + A_Z^2}$ (three features).
- Number of peaks, elapsed time between consecutive local peaks of accelerometer components $A_X, A_Y, A_Z$ along three axes $X, Y, Z$, as well as the variance of the number of peaks of $A_X, A_Y, A_Z$ (seven features).
- Correlations between $A_X$ and $A_Y$, $A_Y$ and $A_Z$, $A_Z$ and $A_X$ (three features).
- Dive phase, encoded by one-hot encoding.

A window was marked as positive if more than 50% of its corresponding output values belonged to a buzz, and negative otherwise (Figure 5). We had 6,348 positive windows out of 526,086 windows against 27 features, enough for robust maximum likelihood estimation of the logistic regression model following the *one in ten* rule (Peduzzi, et al., 2006; Harrell, 2015). We also tried to test whether logistic regression worked better when detecting only the start of buzzes instead of the whole length of the buzzes, however, the performance was worse (results not shown).

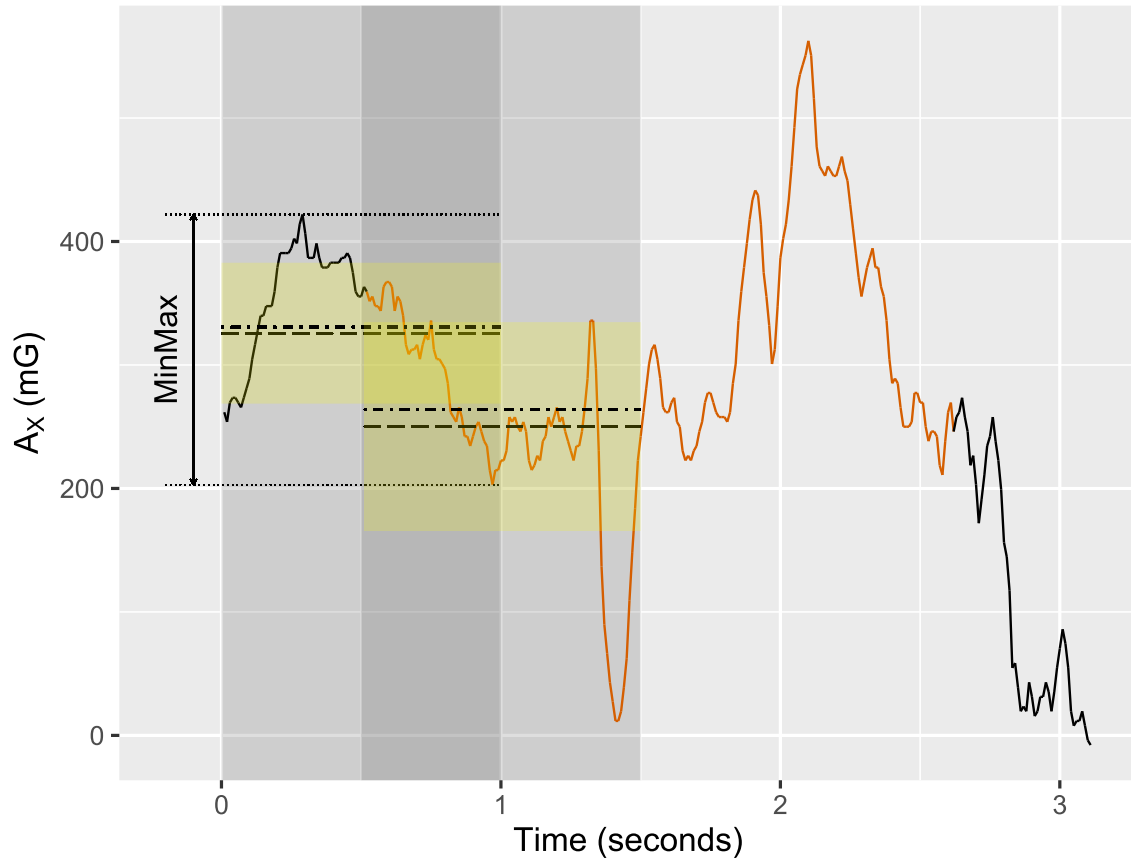

*Figure 5: Illustration of feature extraction from the accelerometer of the X-axis. The dark shaded area indicates where two windows of size of 1-second have 50% overlap. Orange part of the curve indicates the duration of a buzz. Dashed lines are the means of each window, while shaded yellow areas indicate +/- one standard deviation. Dot-dashed lines indicate RMS's of each window.*

## Results

*Machine learning models*

Model hyperparameters of the U-Net models were tuned using a smaller dataset from a single narwhal 168433. The best U-Net models were those with four hidden units. The features of the data for the U-Net model were accelerometer components $A_X, A_Y, A_Z$ and the depth data. The model for all five whales having the best validation loss is presented in Figure 6. Dice loss was smallest at the 224th epoch of the validation set, at which the parameters were chosen for U-Net models.

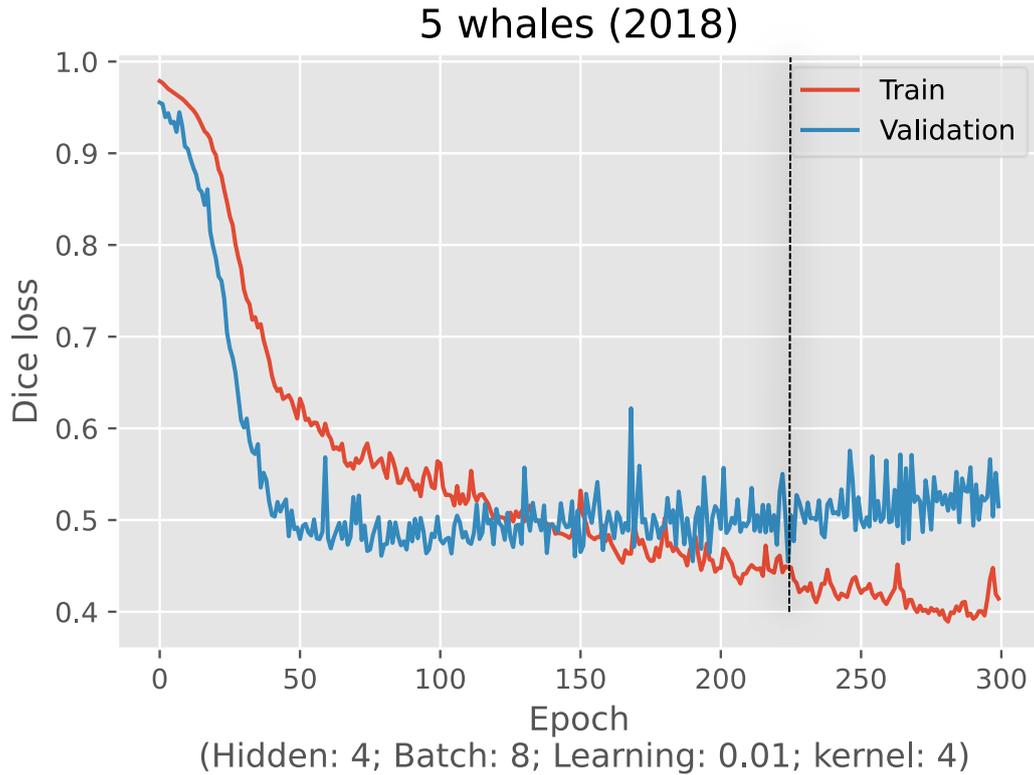

*Figure 6: The training and validation loss when the number of hidden units at first convolution layer are four of the best models with respect to hyper-parameters tuning*

We evaluated the U-Net models on the five test sets from each whale as well as the entire test set in Figure 7. It shows the proportion of correct predictions of buzzes of the models, where we define a correct prediction with some slack: the percentage overlap between predicted and true buzzes. An example for the cases of 50% percentage overlap and 1 second distance between predicted and true buzzes is shown in Figure 8. The proportion of correct predictions of buzzes with a maximum distance to its nearest true buzz smaller than 0.1 seconds and 0.5 seconds (only for U-Net, since the other two models, random forest and logistic regression, only have a resolution of one second), and 1 second, 2 seconds, 3 seconds, 4 seconds, 5 seconds were calculated for all the models (Figure 7).

The U-Net with cross validation performed similarly on each whale, except for narwhal 20158, compared to the U-Net trained and validated on data from all whales. This is reassuring since the algorithm then could possibly generalize well to the narwhal population. There were almost no differences between 500, 1000, and 2000 trees for random forest (results not shown), so we chose the one with 2000 trees. This random forest model predicted poorly on the raw data and even worse with lowpass filtered data of 0.25 Hz (results not shown). Finally, logistic regression models predicted better than random forest, but not as good as the U-Net.

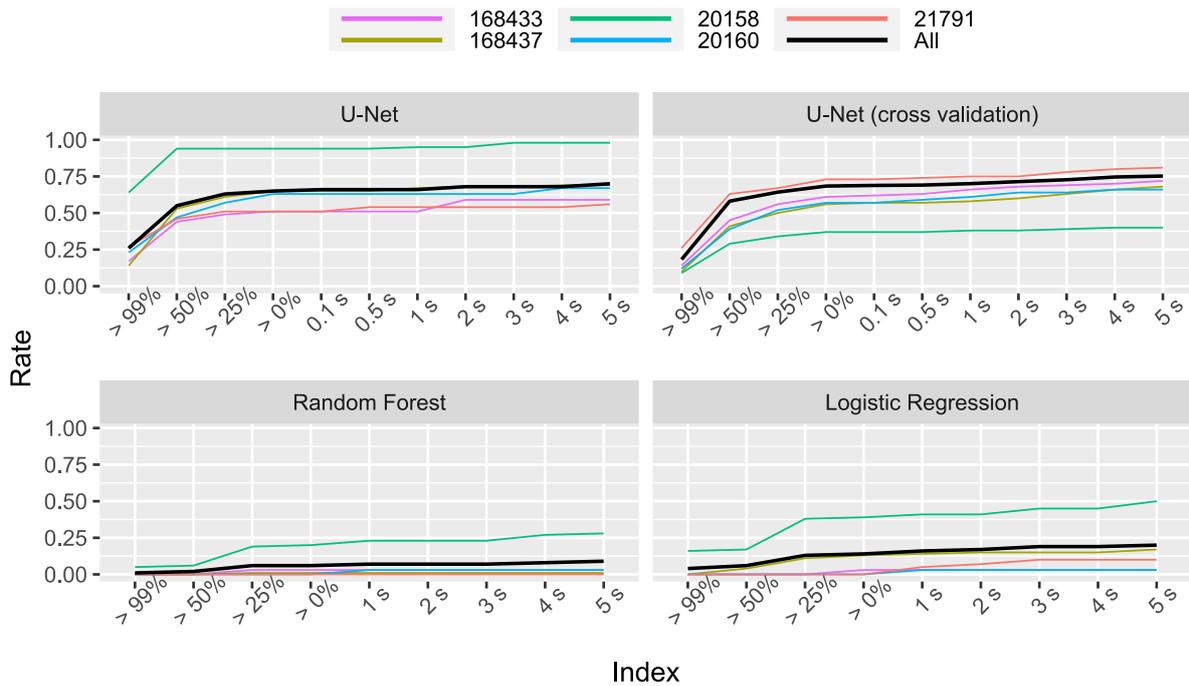

*Figure 7: The proportion of correct predictions of buzzes of the four models: U-Net, U-Net (cross validation), random forest, and logistic regression. On X-axis is shown: the first four indices of the overlap between predicted and true buzzes; the next indices show the proportion of predicted buzzes with a maximum distance to its nearest true buzz smaller than 0.1 seconds, 0.5 seconds (only for U-Net), and 1 second, 2 second, 3 second, 4 seconds and 5 seconds for all the models. The colored lines are the values for each whale, the black line is for all five whales together.*

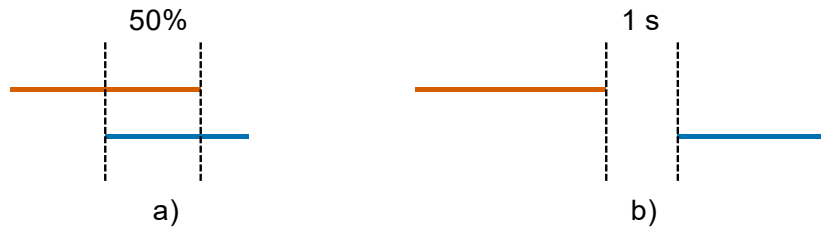

*Figure 8: Examples of the definition of partial correct prediction: a) 50% overlap, and b) distance of one second between ground truth (orange) and prediction (blue)*

If the focus of the analysis is to identify foraging dives, i.e., those dives where buzzes occur, and in that case, how many buzzes the whale emits during that dive, the results improved the classification of dives with feeding activities. We evaluated whether the methods could distinguish between foraging dives with buzzes and exploring dives without buzzes, as well as whether the number of buzzes in each dive could be predicted, even if the exact timing of the buzzes were wrong. Technically, we evaluated the models by counting the number of consecutive series of ones (i.e., buzzing events). There were 456 dives in total, among them 152 were foraging dives, i.e., having buzzes (33.3%). The number of predicted buzzes against the number of true buzzes within each dive is plotted in Figure 9, for U-Net, random forest and logistic regression. Only the U-Net model distinguished well between dives with buzzes and those without buzzes. For the U-Net model, the number of true negatives were 39, false positives were 3, false negatives were 0, and true positives were 23, thus, for identifying foraging/non-foraging dives, the precision was 88% and the recall was 100%. Furthermore, it captured the trend of the number of buzzes within dives well, even if slightly overestimated. The random forest model correctly classified the non-foraging dives, however, for foraging dives it always underestimated the number of buzzes, often as zero. Logistic regression predicted slightly better, but it also underestimated the number of buzzes in each dive and estimated too many non-foraging dives.

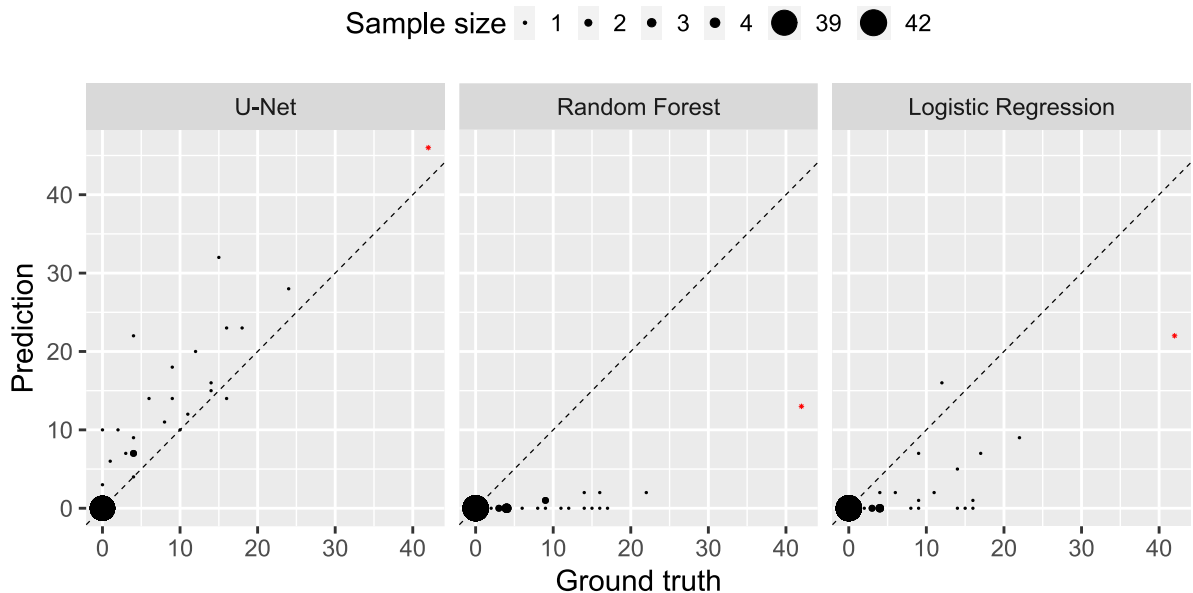

Figure 9: Scatter plot of the number of predicted buzzes against the number of true buzzes per dive for U-Net, random forest, and logistic regression. Size of the dots indicate the number of points. The dashed line is the identity line. The red points indicate the dive illustrated in Figure 11.

In Figure 10, we compared the differences of the number of buzzes and of the time spent buzzing per dive. The U-Net model tended to predict more buzzes than the ground truth, while random forest and logistic regression models predicted less than the ground truth, as also shown in Figure 11. The same pattern emerged for the buzz lengths. All in all, the U-Net performed best on both predictions of number of buzzes as well as on the time spent buzzing per dive (length of buzzing period).

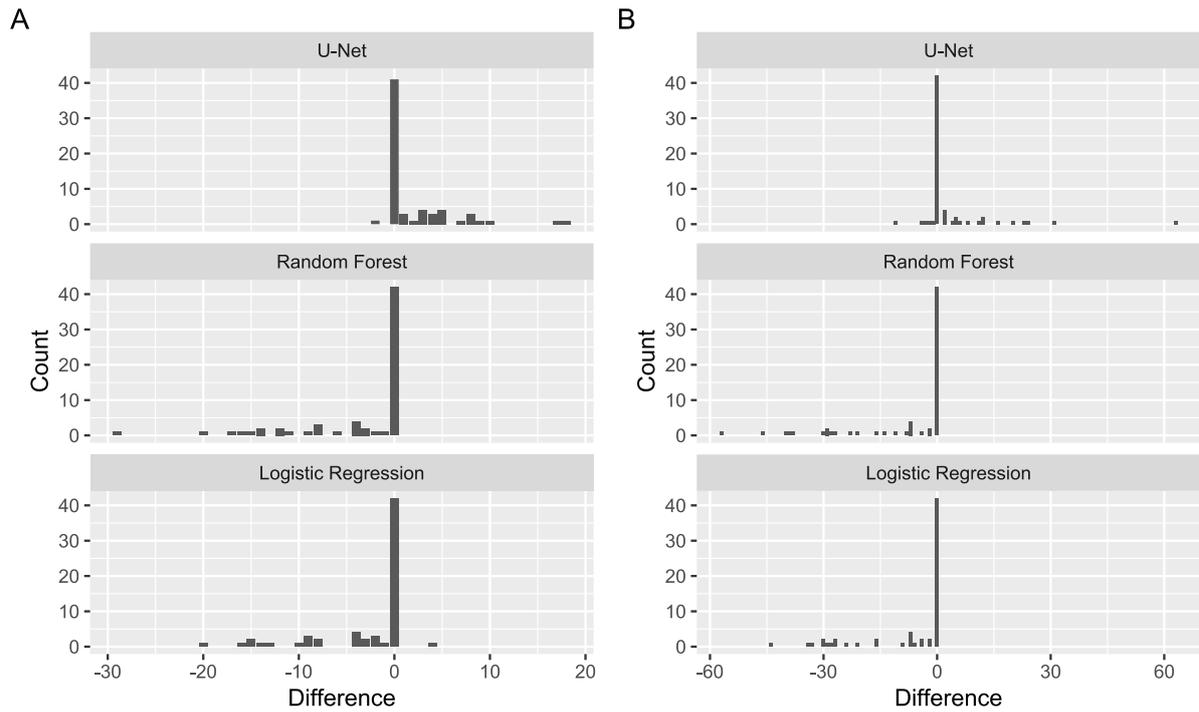

*Figure 10: Histograms of A) the difference between the number of buzzes from the predictions and the ground truth, and B) the difference between the sum of the lengths of the buzzes per dive for U-Net, random forest, and logistic regression*

An example of the predictions compared to the ground truth from the three models is shown in Figure 11. It illustrates the dive with most buzzes, which is a dive from narwhal 20158, showing clearly, that the U-Net model predicted best with many overlaps between predictions and ground truth, while logistic regression came second.

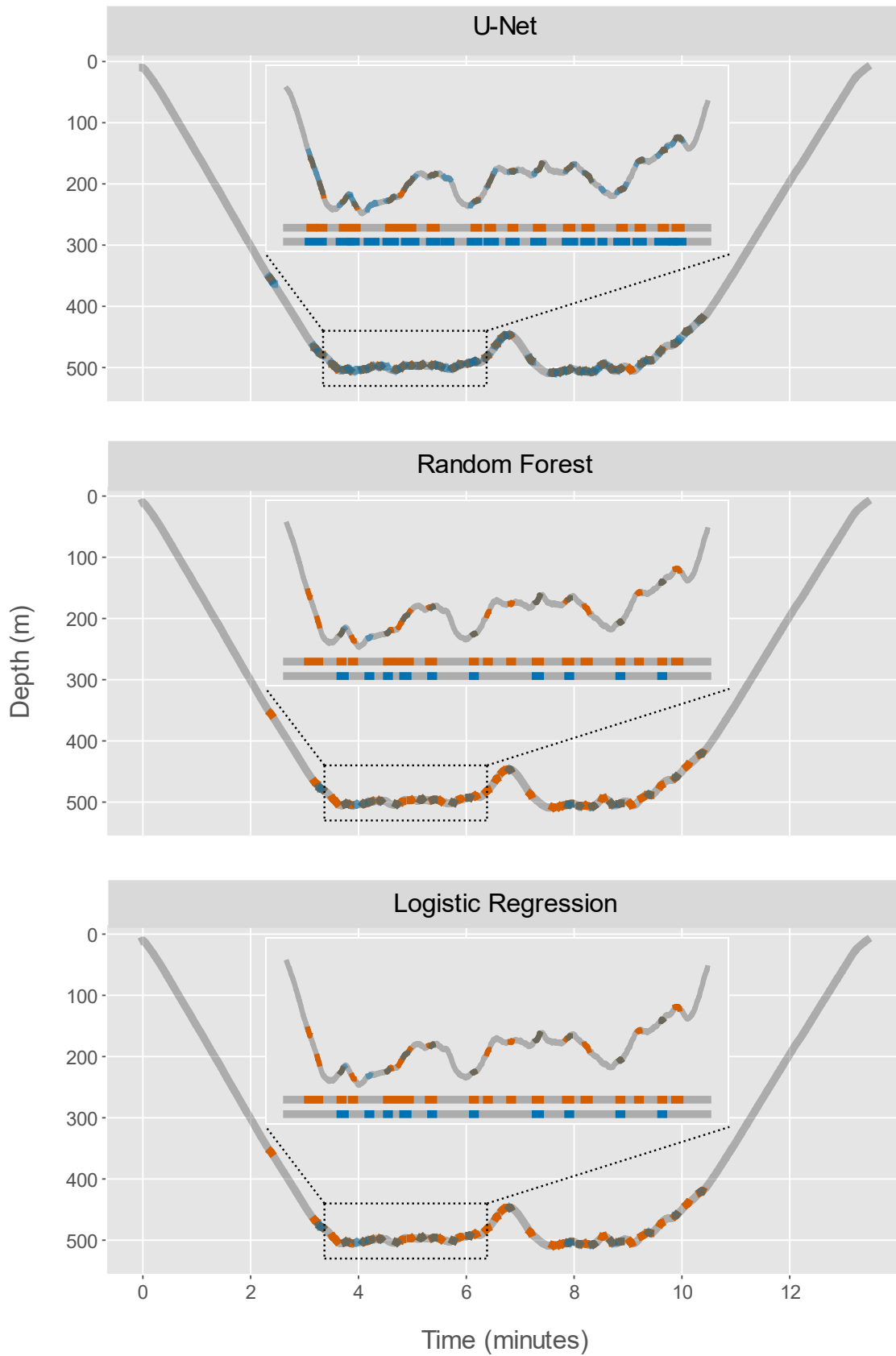

*Figure 11: Example of a dive of the record of narwhal 20158 with ground truth buzz (orange) and prediction buzz (blue). The first panel illustrates the U-Net model, the second illustrates the random forest, and the third illustrates logistic regression. The smaller panel within each panel show a zoomed-in section of the time-depth series marked with a dashed line.*

*Jerk analysis*

Figure 12 shows the precision and recall of RMS jerks at different thresholds. The precision of prediction of buzzes from big RMS jerks is low, less than 0.25, for thresholds less than 12,500 mG/s. It increases for larger thresholds; the precision for narwhals 168437 and 168433 even reach 1 for some thresholds, but the true positives and the recalls decrease extremely fast to close to zero. Note that for a threshold of zero, the precision equals the proportion of ones in the data, and the recall equals one. Additional attempts with a delayed jerk within 1 second (0.2, 0.4, 0.6, 0.8, and 1 second) after the buzzes can be found in Supplementary Material. Figure 13 shows an example trace, with several high RMS peaks without any buzz activity, while there are a few high RMS peaks close to the buzzes. We therefore conclude that jerks are not a suitable criterion for detecting buzzes and prey capture events.

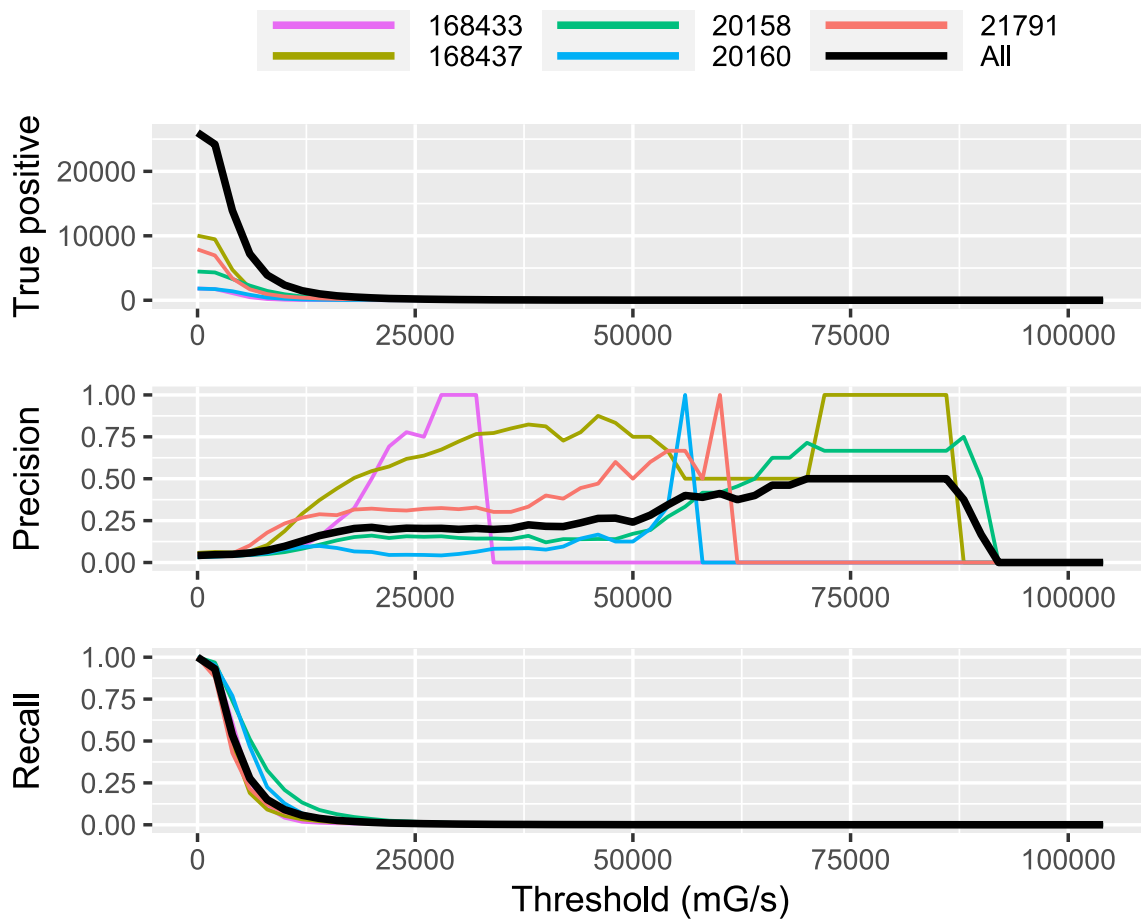

*Figure 12: Precision and recall of RMS jerks for different thresholds for predicting buzzes.*

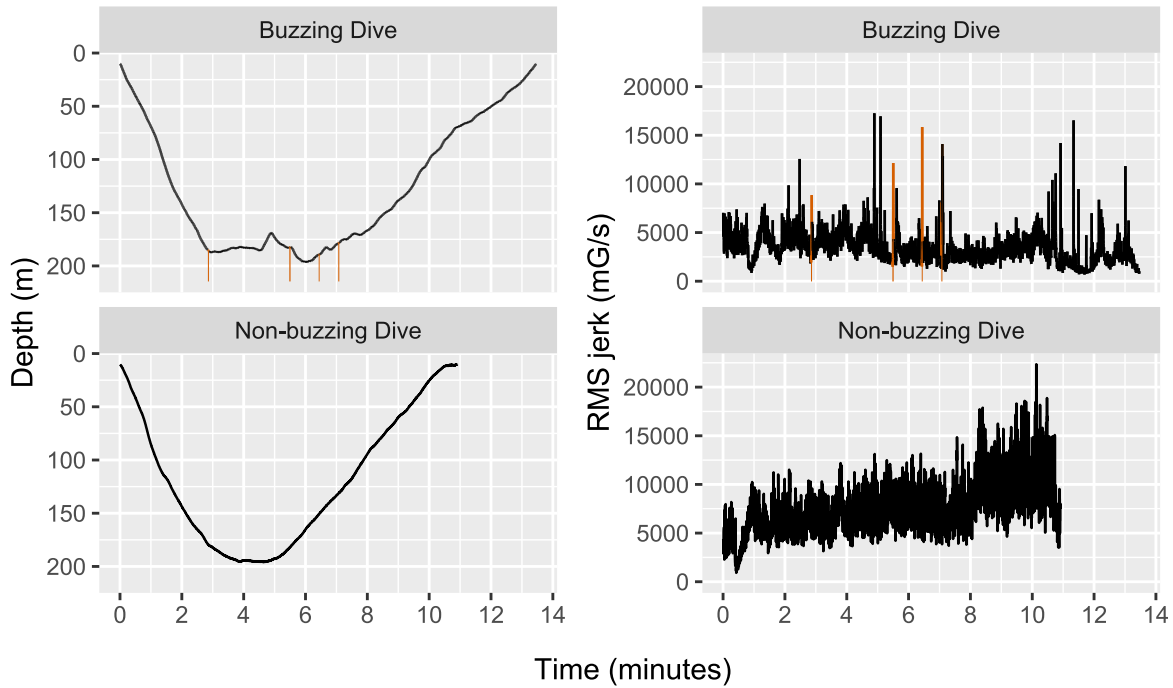

*Figure 13: Example of two dives of the record of narwhal 21791. The upper two panels show the time-depth series and RMS jerk of a buzzing dive; the lower two panels show the time-depth series and RMS jerks of a non-buzzing dive. The orange lines indicate presence of buzzes, in the upper right panel the RMS jerks are colored orange when buzzing.*

## Conclusions

In this study, we investigated if some special movement patterns present around the times of buzzes from free ranging narwhals can be detected by machine learning methods. Our results show that the U-Net can be used to detect buzzes from accelerometer data. We also examined whether the narwhals make big jerks around buzzes, which have been found in a previous study of captive harbor seals (Ydesen, et al., 2014). They used a triaxial accelerometer to collect head- and jaw mounted accelerometer data in prey capturing attempts. It works well also in a wild environment for harbor porpoises (Wisniewska, et al., 2016), as well as sperm whales (Fais, et al., 2016). In our study, the tags are positioned on the back of narwhals, so it may not detect more subtle head- and jaw-jerks, but major body movements towards targeted prey during buzzing events should be detectable. We frequently identified acceleration peaks in the narwhal data that did not follow a buzzing event, and analysis showed that both the precision and recall were poor. Moreover, we tested on free ranging narwhals rather than captive specimens, so the variances are larger. The narwhals may engage in many different movement activities that imply quick movements, so false positives might be high if only big RMS jerks are used as a criterion. We therefore conclude that big RMS jerks are not trustworthy indicators for detection of buzz events. From an anatomical perspective the absence of teeth in the jaws also makes raptorial feeding less likely and suggests that narwhals ingest prey by buccal suction feeding. The narwhals that were instrumented for this study feed on squids in the water column that presumably are easy to capture and ingest. Other narwhal populations feed on halibut that may require a more raptorial capturing approach and rapid movements during the buzz phase of the prey strike.

With the improvement of tagging technologies, more data especially from accelerometer instruments can be collected and tools like machine learning for big data analysis might contribute enormously to the understanding of marine predators. We have demonstrated an application of deep learning, with U-Net, to accelerometer and depth data for detection of buzzes in narwhals. The performance of U-Net was superior to random forest, the baseline method of tabular dataset, which failed to detect the buzzes. We used the Dice loss function, which is suitable for an imbalanced dataset. The trained model can be used to make predictions or facilitate the training process on new datasets, called transfer learning (Pratt, 1993). It distinguished well between foraging dives with buzzes and exploring dives without buzzes, much better than random forest and logistic regression. Its buzz predictions were much closer to the ground truth than the two other models' predictions as well.

Finding the right features for random forest or logistic regression is particularly hard in new applications. Thus, we cannot definitively conclude that U-Net, or more general deep learning, is superior without further research. A simple method like logistic regression performed better than random forest, although worse than U-Net. Furthermore, the determination of the right hyper-parameters for the U-Net is computationally expensive. The performance of logistic regression might be improved by more careful feature selection such as including correlation of buzzes. Logistic regressions have an advantage of being much simpler and much more transparent than the U-Net or deep learning in general.

Although our study shows positive results on the use of U-Net models, there are several limitations that require more analysis. For example, deep learning methods, in general, are not transparent because of the huge number of parameters. The general way of learning is by trial-and-error, i.e., to test different hyper-parameters and/or loss functions. It creates an enormous training time, as well as a high carbon footprint (Anthony, et al., 2020), and makes it vulnerable to spurious findings due to the lack of transparency. Combining signal processing techniques and more transparent statistical/machine learning methods, may help to understand when and why the methods work (Forde & Paganini, 2019; Succi & Coveney, 2019). However, our results provide some evidence that deep learning provides a valuable tool compared to traditional machine learning or statistical methods. The supervised machine learning approaches in this study could be extended to any other marine mammals' datasets.

## Acknowledgments

We would like to thank Susanna Blackwell and Alexander Conrad for supporting of data labelling of narwhal buzzing activity. SD was supported by Independent Research Fund Denmark, case: 9040-00215B.20 of 28

# References

Alsheikh, M. A. et al., 2015. Deep Activity Recognition Models with Triaxial Accelerometers. *The Workshops of the Thirtieth AAAI Conference on Artificial Intelligence.*

Anthony, L. F. W., Kanding, B. & Selvan, R., 2020. Carbontracker: Tracking and Predicting the Carbon Footprint of Training Deep Learning Models. *ICML Workshop on Challenges in Deploying and monitoring Machine Learning Systems.*

Bayat, A., Pomplun, M. & Tran, D. A., 2014. A Study on Human Activity Recognition Using Accelerometer Data from Smartphones. *Procedia Computer Science,* pp. 450-457.

Berta, A., Sumich, J. L. & Kovacs, K. M., 2015. Sound Production for Communication, Echolocation, and Prey Capture. I: *Marine Mammals: Evolutionary Biology, Third Edition.* s.l.:Academic Press, pp. 345-395.

Blackwell, S. et al., 2018. Spatial and temporal patterns of sound production in East Greenland narwhals. *PLoS ONE.*

Boehmke, B. & Greenwell, B. M., 2020. *Hands-On Machine Learning with R.* s.l.:Chapman and Hall/CRC.

Breiman, L., 2001. Random Forests. *Machine Learning,* pp. 5-32.

Chen, C., Liaw, A. & Breiman, L., 2004. *Using random forest to learn,* Berkeley: University of California.

Drozdzal, M. et al., 2016. The Importance of Skip Connections in Biomedical Image Segmentation. *International Workshop on Deep Learning in Medical Image Analysis,* pp. 179-187.

Fais, A. et al., 2016. Sperm whale predator-prey interactions involve chasing and buzzing, but no acoustic stunning. *Scientific Reports.*

Farabet, C., Couprie, C., Najman, L. & LeCun, Y., 2013. Learning hierarchical features for scene labeling. *IEEE Trans Pattern Anal Mach Intell.,* pp. 1915-1929.

Forde, J. Z. & Paganini, M., 2019. The Scientific Method in the Science of Machine Learning. *ArXiv.*

Goodfellow, I., Bengio, Y. & Courville, A., 2016. *Deep Learning.* s.l.:MIT Press.

Google, 2020. *Colaboratory.* [Online]
Available at: https://research.google.com/colaboratory/faq.html

Graham, Z. A., Garde, E., Heide-Jørgensen, M. P. & Palaoro, A. V., 2020. The longer the better: evidence that narwhal tusks are sexually selected. *Biology Letters.*

Harrell, F., 2015. *Regression Modeling Strategies: With Applications to Linear Models, Logistic and Ordinal Regression, and Survival Analysis.* s.l.:Springer International Publishing.

Heide-Jørgensen, 2009. Narwhal Monodon monoceros. I: *Encyclopedia of Marine Mammals, 2nd Edition.* s.l.:s.n., p. 754–758.

Heide-Jørgensen, M., Dietz, R. & Leatherwood, S., 1994. A note on the diet of narwhals (Monodon monoceros) in Inglefield Bredning (NW Greenland). *Biosci.,* p. 213–216.

Heide-Jørgensen, M. et al., (In prep). Effects of seismic air gun pulses on narwhals. *Frontiers in Marine Science.*

Heide-Jørgensen, M. et al., 2015. The predictable narwhal: satellite tracking shows behavioural similarities between isolated subpopulations. *Journal of Zoology 297,* pp. 54-65.

He, K., Zhang, X., Ren, S. & Sun, J., 2016. Deep Residual Learning for Image Recognition. *IEEE Conference on Computer Vision and Pattern Recognition (CVPR),* pp. 770-778.

Hillman, G. et al., 2003. Computer-assisted photo-identification of individual marine vertebrates: a multi-species system. *Aquatic Mammals,* pp. 117-123.

Ho, T. K., 1995. Random Decision Forests. *Proceedings of the 3rd International Conference on Document Analysis and Recognition,* pp. 278-282.

Kingma, D. & Ba, J., 2015. Adam: A method for stochastic optimization. *International Conference on Learning Representations,* pp. 1-15.

Krizhevsky, A., Sutskever, I. & Hinton, G., 2012. Imagenet classification with deep convolutional neural networks. *Advances in Neural Information Processing Systems,* p. 1097–1105.

Kwapisz, J. R., Weiss, G. M. & Moore, S. A., 2010. Cell phone-based biometric identification. *Fourth IEEE International Conference on Biometrics: Theory Applicationsand Systems (BTAS).*

Laidre, K. & Heide-Jørgensen, M., 2005. Winter feeding intensity of narwhals (Monodon monoceros). *Marine Mammal Science,* p. 45–57.

Luque, S. P. & Fried, R., 2011. Recursive Filtering for Zero Offset Correction of Diving Depth Time Series with GNU R Package diveMove. *PLoS One.*

Ng, A., 2015. *Deep Learning.* s.l., s.n.

Nowacek, D. P. et al., 2016. Studying cetacean behaviour: new technological approaches and conservation applications. *Animal Behaviour,* pp. 235-244.

Paszke, A. et al., 2017. Automatic differentiation in PyTorch. *NIPS 2017 Autodiff Workshop: The Future of Gradient-based Machine Learning Software and Techniques.*

Pedregosa, F. et al., 2011. Scikit-learn: Machine Learning in Python. *Journal of Machine Learning Research,* pp. 2825-2830.

Peduzzi, P. et al., 2006. A simulation study of the number of events per variable in logistic regression analysis. *Journal of Clinical Epidemiology,* pp. 1373-1379.

Perslev, M. et al., 2019. U-Time: A Fully Convolutional Network for Time Series Segmentation Applied to Sleep Staging. *Advances in Neural Information Processing Systems (NeurIPS).*

Pratt, L., 1993. Discriminability-based transfer between neural networks. *NIPS Conference: Advances in Neural Information Processing Systems,* pp. 204-211.

Probst, P., Boulesteix, A.-L. & Bischl, B., 2019. Tunability: Importance of Hyperparameters of Machine Learning Algorithms. *Journal of Machine Learning Research 20,* pp. 1-32.

Ronneberger, O., Fischer, P. & Brox, T., 2015. U-net: Convolutional networks for biomedical image segmentation. *International conference on medical image computing and computer-assisted intervention,* pp. 234-241.

Shepard, E. et al., 2008. Identification of animal movement patterns using tri-axial accelerometry. *Endanger. Species Res.,* pp. 47-60.

Smith, A. G., Petersen, J., Selvan, R. & Rasmussen, C. R., 2020. Segmentation of roots in soil with U-Net. *Plant Methods.*

Succi, S. & Coveney, P. V., 2019. Big data: the end of the scientific method?. *Philosophical transaction of the royal society A.*

Swanson, D. C., 2008. Acoustic Data Acquisition. I: *Handbook of Signal Processing in Acoustics.* New York, NY: Springer New York, pp. 17-32.

Szegedy, C. et al., 2015. Going deeper with convolutions. *Proceedings of the IEEE conference on computer vision and pattern recognition,* pp. 1-9.

Tervo, O. M. et al., 2020. Hunting by the stroke: How foraging drives diving behavior and swimming biomechanics of East-Greenland narwhals (Monodon monoceros). *Submitted.*

Tompson, J. et al., 2015. Efficient object localization using convolutional networks. *Proceedings of the IEEE conference on computer vision and pattern recognition,* pp. 648-656.

Visa, S. & Ralescu, A., 2005. Issues in mining imbalanced data sets-a review paper. *Proceedings of the sixteen midwest artificial intelligence and cognitive science conference,* pp. 67-73.

Wang, G., 2019. Machine learning for inferring animal behavior from location and movement data. *Ecological Informatics,* pp. 69-76.

Warmerdam, V. D., 2018. *Winning with Simple, even Linear, Models,* London: PyData.

Williams, R. J., Hinton, G. E. & Rumelhart, D. E., 1986. Learning representations by back-propagating errors. *Nature 323 (6088),* pp. 533-536.

Wilson, R., Shepard, E. & Liebsch, N., 2008. Prying into the intimate details of animal lives: use of a daily diary on animals. *Endanger. Species Res.,* pp. 123-137.

Wisniewska, D. M. et al., 2016. Ultra-High Foraging Rates of Harbor Porpoises Make Them Vulnerable to Anthropogenic Disturbance. *Current Biology,* pp. 1441-1446.

Ydesen, K. et al., 2014. What a jerk: prey engulfment revealed by high-rate, super-cranial accelerometry on a harbour seal (Phoca vitulina). *The Journal of Experimental Biology,* pp. 2239-2243.

# Supplementary material

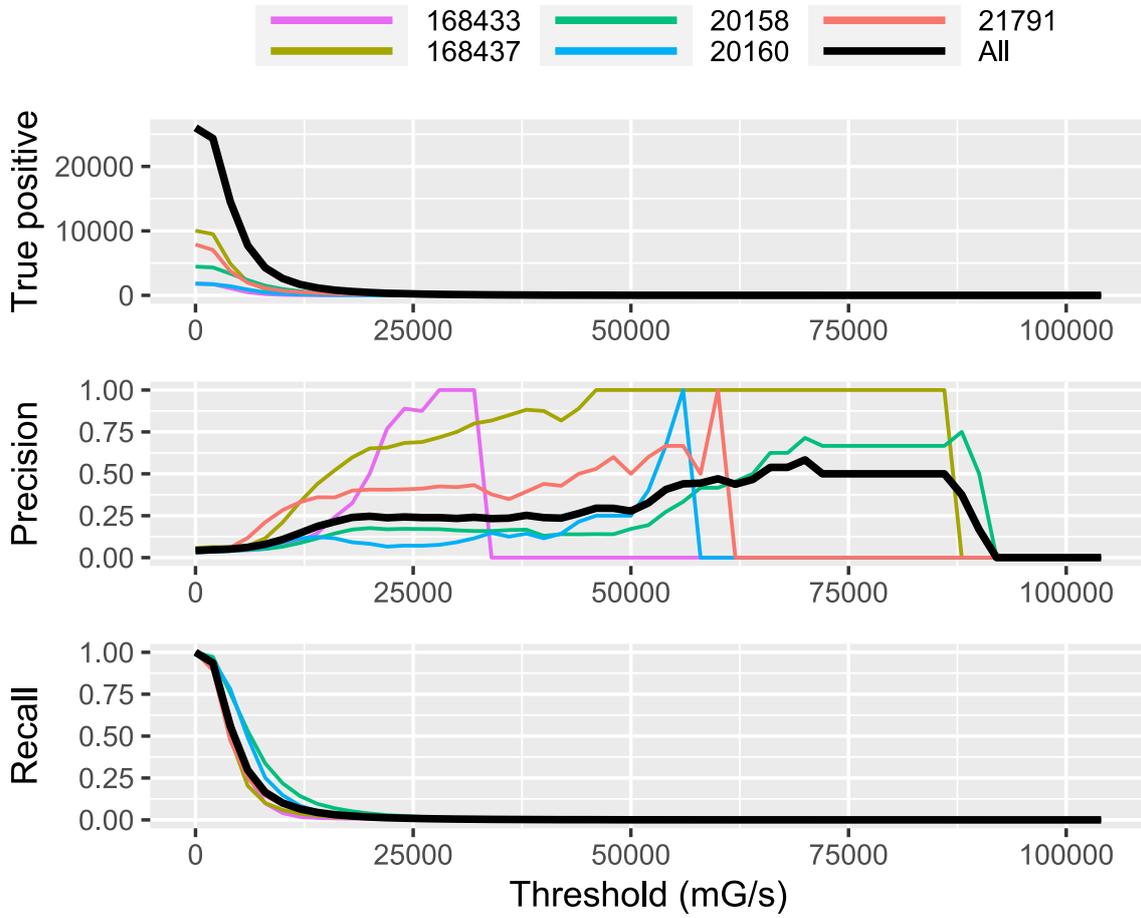

*a) 0.2 seconds*

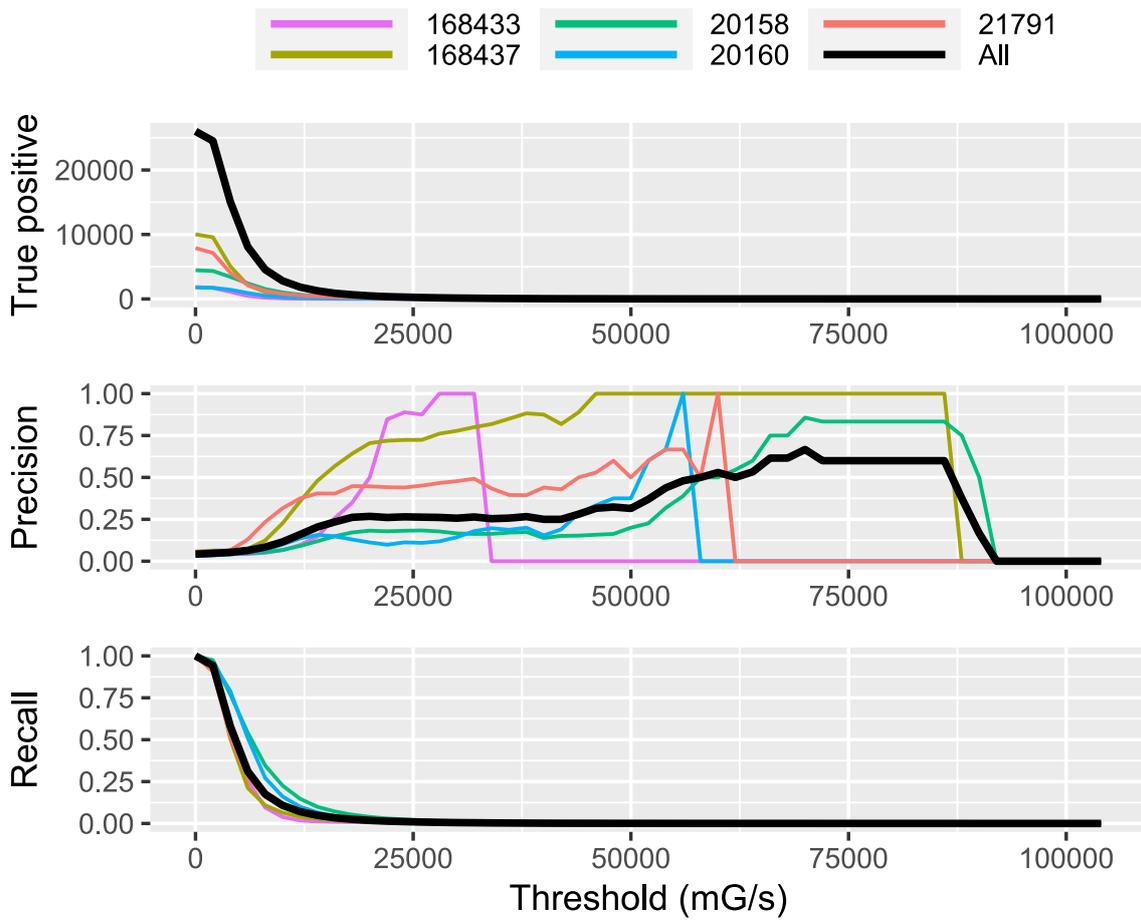

b)  0.4 seconds

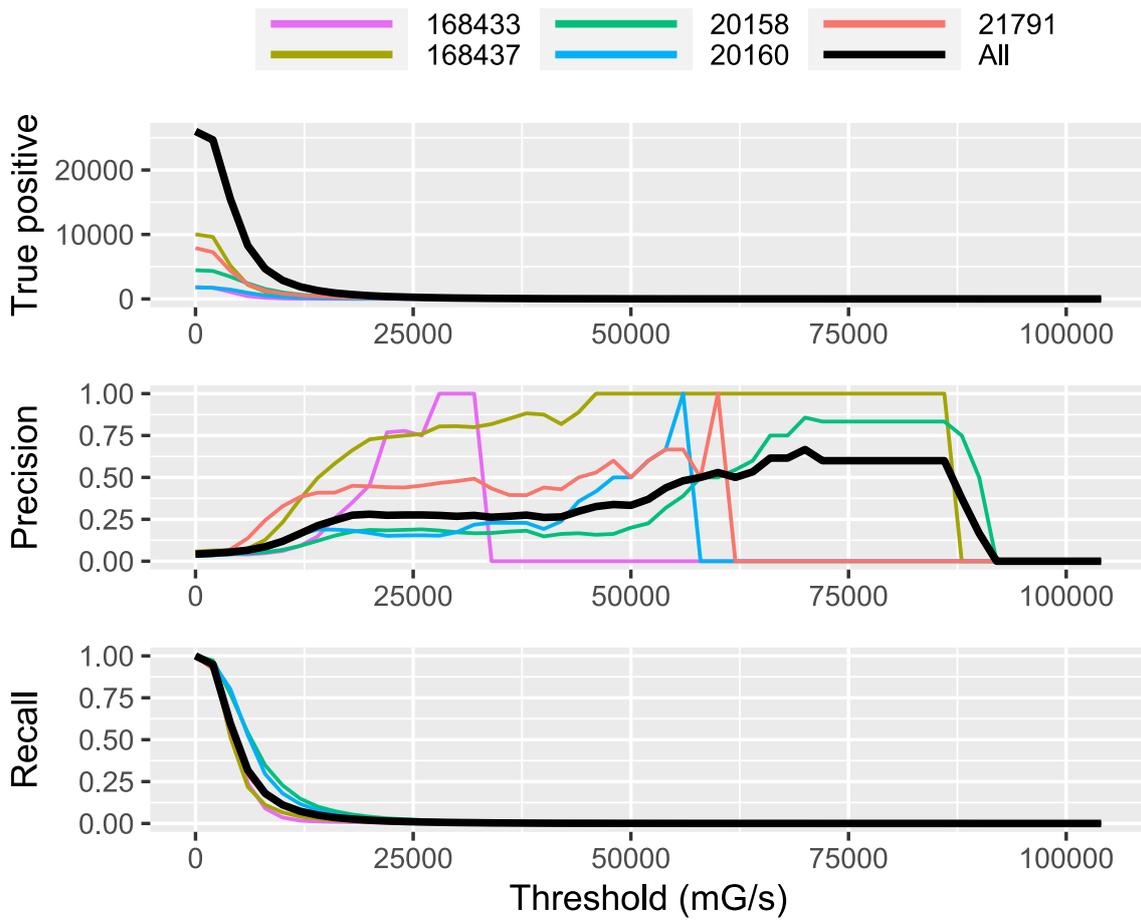

c) *0.6 seconds*

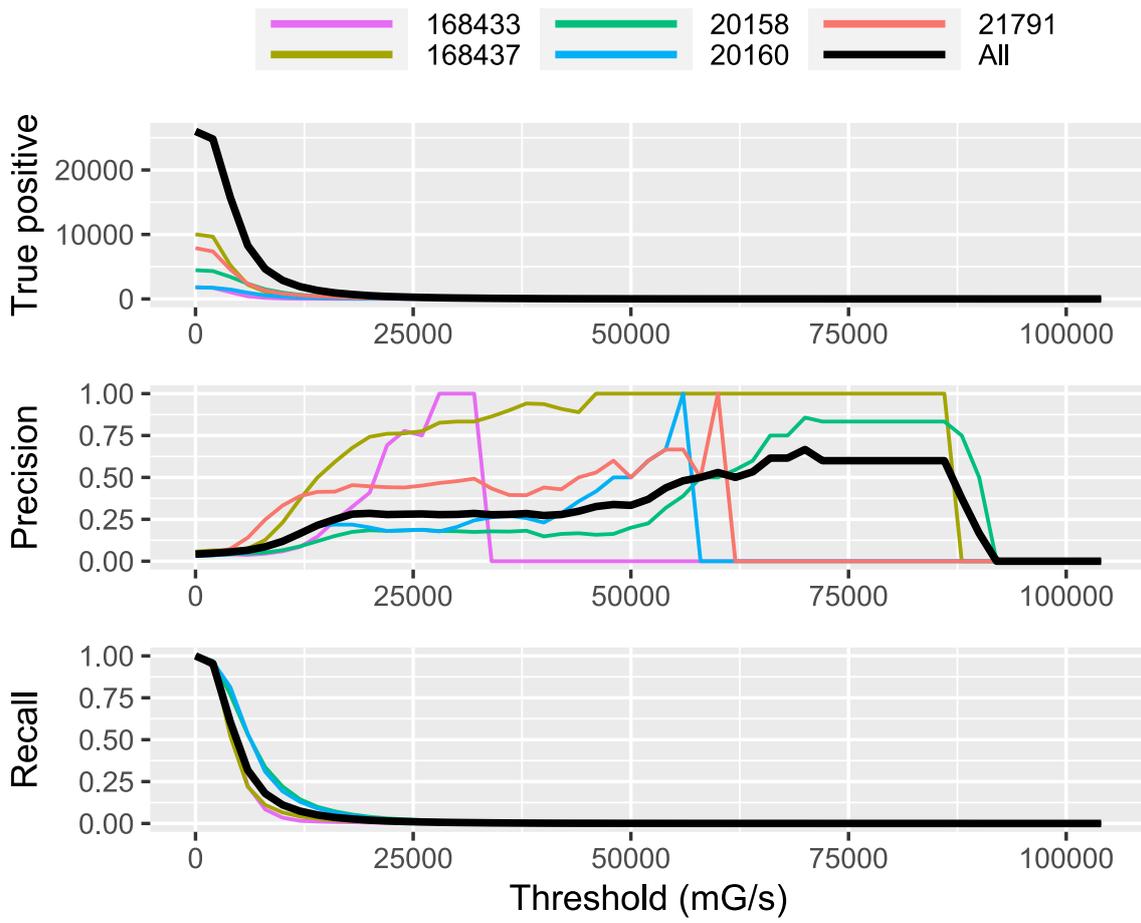

d) 0.8 seconds

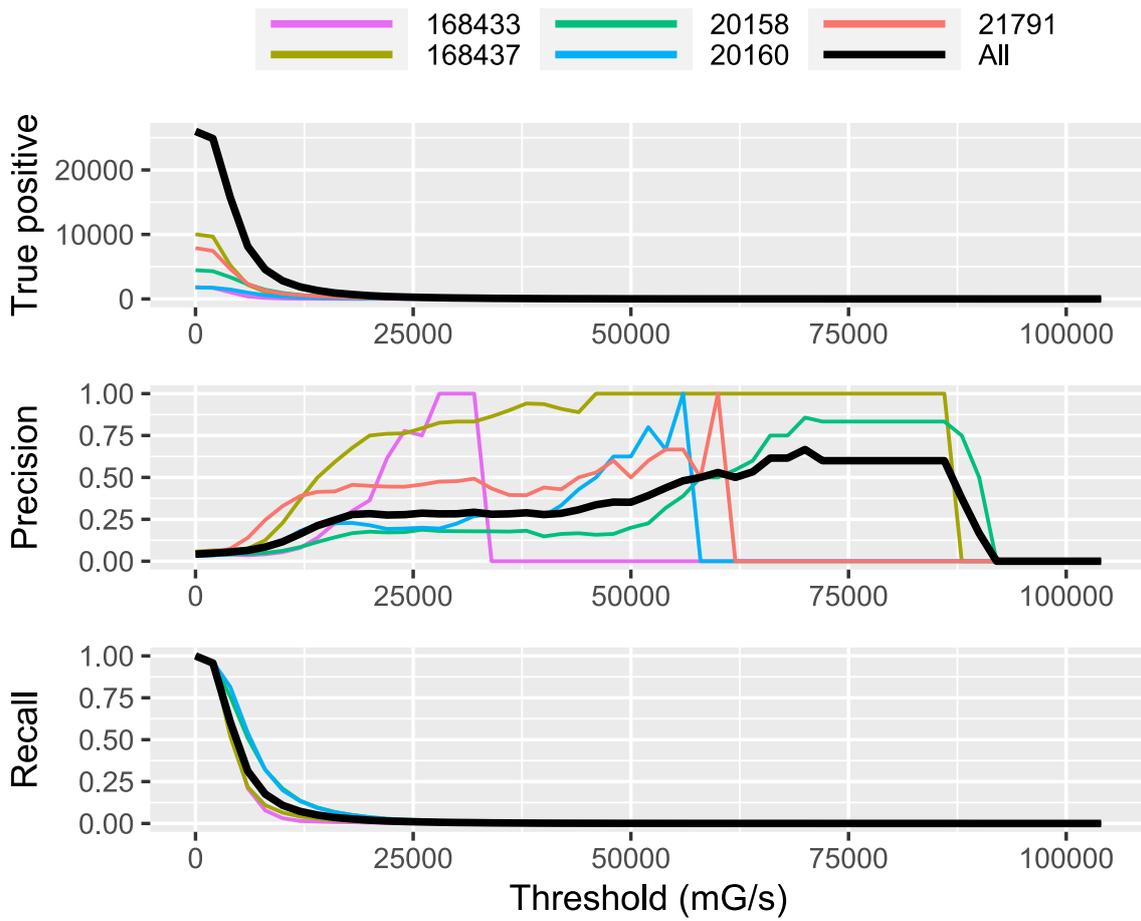

*e) 1 second*

*Figure S1: Precision and recall of RMS jerks for different thresholds for predicting buzzes for different delays of 0.2, 0.4, 0.6, 0.8 and 1 second.*